\documentclass[preprint,12pt,3p]{elsarticle}

\usepackage{caption,subcaption}
\usepackage{float}
\usepackage{ragged2e}
\usepackage{multirow}
\usepackage{enumitem}
\usepackage{enumitem}
\usepackage{amsmath}
\usepackage{graphicx}
\usepackage{booktabs}
\usepackage[table]{xcolor}
\usepackage{courier} 
\usepackage{listings}

\usepackage[
    protrusion=true,
    activate={true,nocompatibility},
    final,
    tracking=true,
    kerning=true,
    spacing=true,
    factor=1100]{microtype}
\SetTracking{encoding={*}, shape=sc}{40}

\definecolor{darkred}{rgb}{0.5,0,0}     
\definecolor{darkgreen}{rgb}{0,0.5,0}   	
\definecolor{darkblue}{rgb}{0,0,0.5}    
\definecolor{airforceblue}{rgb}{0.36, 0.54, 0.66}
\definecolor{arsenic}{rgb}{0.23, 0.27, 0.29}
\definecolor{cadetblue}{rgb}{0.37, 0.62, 0.63}

\newcommand{\change}[1]{{\color{darkgreen}#1}}

\newcommand{\secref}[1]{Section~\ref{sec:#1}}
\newcommand{\tabref}[1]{Table~\ref{tab:#1}}

\newcommand{\figref}[1]{Figure~\ref{fig:#1}}

\usepackage[colorlinks,bookmarksnumbered,linkcolor=darkblue,citecolor=darkred,urlcolor=darkgreen]{hyperref}

\usepackage{pifont}
\newcommand{\cmark}{\ding{51}}%
\newcommand{\xmark}{\ding{55}}%

\usepackage{tikz}
\usepackage{pgfplots}

\newcommand*{\twoelementtable}[3][l]%
{%
	\renewcommand{\arraystretch}{0.8}%
	\begin{tabular}[t]{@{}#1@{}}%
		#2\tabularnewline
		#3%
	\end{tabular}%
}

\journal{Journal of Systems and Software}

\begin{document}

\begin{frontmatter}

\title{Generating Summaries for Methods of Event-Driven Programs: an Android Case Study}


\author[label5]{Alireza~Aghamohammadi\fnref{label3}}
\ead{aaghamohammadi@ce.sharif.edu}

\author[label5]{Maliheh~Izadi\fnref{label3}}
\ead{maliheh.izadi@sharif.edu}

\fntext[label3]{\change{The first two authors contributed equally to this work.}}

\author[label5]{Abbas~Heydarnoori\corref{cor1}}
\ead{heydarnoori@sharif.edu}
\cortext[cor1]{\change{Corresponding author}}
\address[label5]{A. Aghamohammadi, M. Izadi, and A. Heydarnoori are with the Department of Computer Engineering, Sharif University of Technology, Iran.}

\begin{abstract}
\justifying
The lack of proper documentation makes program comprehension a cumbersome process for developers. Source code summarization is one of the existing solutions to this problem. Lots of approaches have been proposed to summarize source code in recent years. A prevalent weakness of these solutions is that they do not pay much attention to interactions among elements of a software. An element is simply a callable code snippet such as a method or even a clickable button. As a result, these approaches cannot be applied to event-driven programs, such as Android applications, because they have specific features such as numerous interactions between their elements. To tackle this problem, we propose a novel approach based on deep neural networks and dynamic call graphs to generate summaries for methods of event-driven programs. First, we collect a set of comment/code pairs from Github and train a deep neural network on the set. Afterward, by exploiting a dynamic call graph, the Pagerank algorithm, and the pre-trained deep neural network, we generate summaries. An empirical evaluation with 14 real-world Android applications and 42 participants indicates 32.3\% BLEU4 which is a definite improvement compared to the existing state-of-the-art techniques. We also assessed the informativeness and naturalness of our generated summaries from developers' perspectives and showed they are sufficiently understandable and informative.
\end{abstract}


\end{frontmatter}


\section{Introduction}\label{sec:introduction}
During the software development life cycle, various reports and documentation such as requirements specification, architecture documents, design documents, bug reports and so forth need to be generated. Unfortunately, lack of direct motivation in software teams results in inadequate, out-of-date and unqualified documentation. This in return makes program comprehension a difficult and time-consuming task for other team members. Xia et al.~\cite{IEEExia2017measuring} attest to this fact by claiming that on average developers spend about 58\% of their time understanding a program~\cite{IEEExia2017measuring}. Source code summarization aids developers in understanding how a program works better and faster. This technique is used for describing the goal or functionality of different parts of a software program, namely methods, classes, or packages in a comment~\cite{IEEEBadihi2017}. \figref{Example-Source-Code-Summarization} depicts a code snippet with its accompanied comment. 

\begin{figure}[h]
	\centering
	\footnotesize
\begin{lstlisting}[language = Java , frame = trBL , firstnumber = last , escapeinside={(*@}{@*)}]
// Creates an intent, adds location data to it
// as an extra, and starts the intent service
// for fetching address.
private void startIntentService() {
    Intent intent = new Intent(this, FetchAddressIntentService.class);
    intent.putExtra(Constants.RECEIVER, mResultReceiver);
    intent.putExtra(Constants.LOCATION_DATA_EXTRA, mLastLocation);
    startService(intent);}
\end{lstlisting}
			
	\caption{An example code snippet with its summary~\cite{googlesample}}\label{fig:Example-Source-Code-Summarization}
\end{figure}

\begin{figure}[t]
 	\centering
 	\includegraphics[height=8cm]{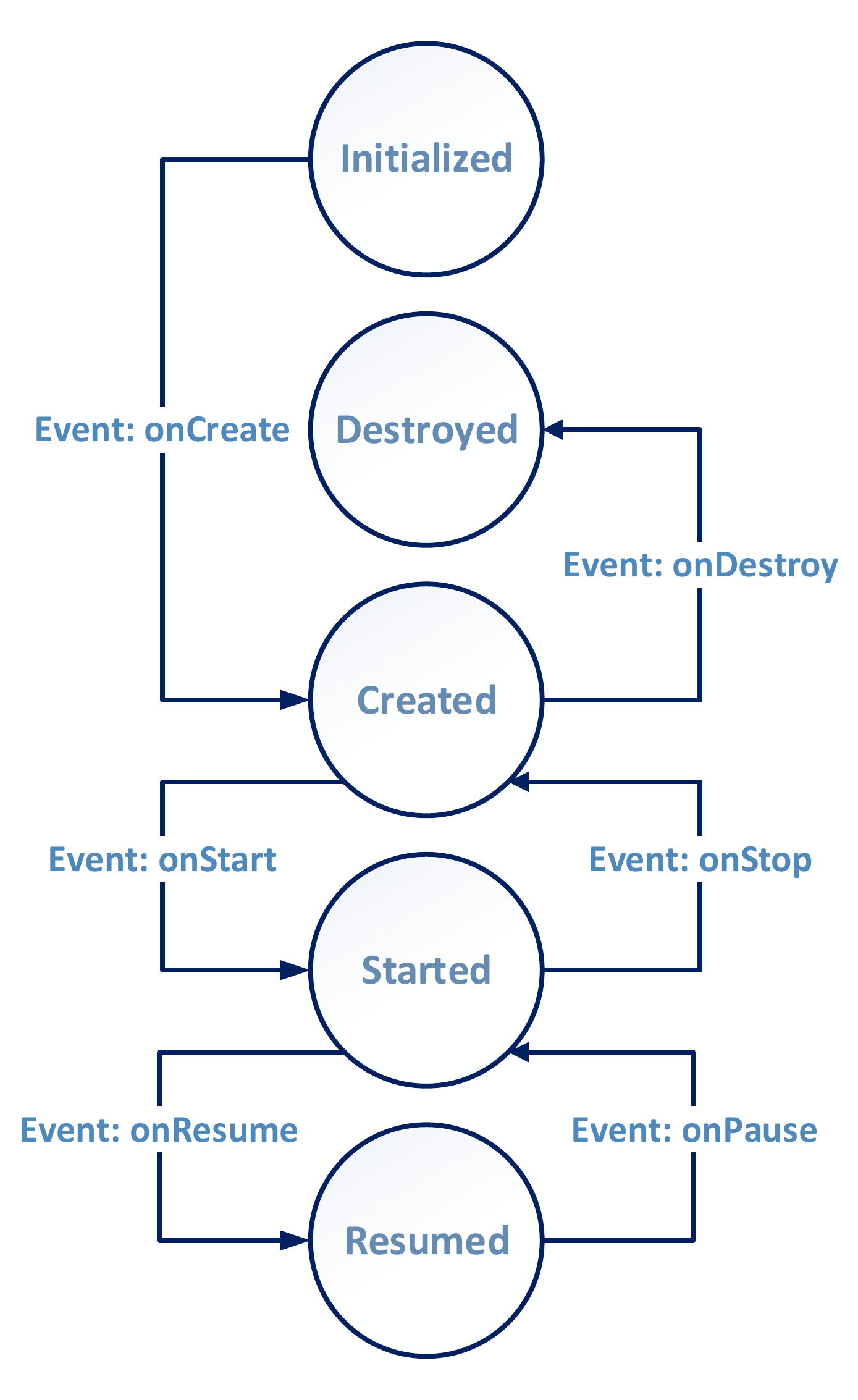}
 	\caption{An Android application's life cycle}\label{fig:Android-LifeCycle}
 \end{figure}

Many approaches for source code summarization have been proposed over the course of past years. For instance, exploiting knowledge of the crowd~\cite{IEEEBadihi2017,FCSNazarJGZLR2016,ICSEGuerroujBR2015,SCAMRahmanRK2015,SANERWongLT2015}, information retrieval~\cite{ASESridharaHMPV2010,IEEEAntoniolCCLM2002,ICPCMorenoASMPV2013}, machine learning~\cite{ICPCMcBurneyLMW2014,IEEEHaiducAMM2010,ICSEHaiducAM2010,IEEEFowkesCRALS2017,ICPCEddyRKC2013},  neural networks~\cite{ICMLAllamanisPS16,Iyer-2016,ICPChu2018deep}, or even tracking eye-movements of developers~\cite{ICSERodeghero2014} are among the approaches for addressing this issue. The main limitation of these approaches is that they do not pay much attention to interactions among elements of a software while generating summaries. McBurney and McMillan~\cite{IEEEMcBurney2016} regarded the context of a program as a critical factor. They defined a method based on its invocations. However, they did not consider events in a software program. Event-driven programs contain a cycle which waits for events. When an event triggers, the program runs it. Therefore, interactions among elements are specified at run-time in these programs.
An Android application is an excellent example of an event-driven program. 
As a result, these approaches cannot be applied to event-driven programs because of characteristics of those programs such as numerous interactions between their elements.
\figref{Android-LifeCycle} demonstrates the life cycle of Android applications.

Consider a program presenting one \texttt{Button} and one \texttt{TextBox} in the main display page. A user types an arbitrary text and pushes the \texttt{Button} which leads to another page. The new page shows the user's text. \figref{Motivation-Example} is a source code that demonstrates the behavior of this program. When a user pushes the \texttt{Button}, the functions \texttt{sendMessage()} from the \texttt{MainActivity} class and \texttt{onCreate()} from the \texttt{DisplayMessageActivity} class are invoked, respectively.
Unlike how trivial it may seem, this is not an easy task for a Java code. Indeed, when a user pushes the \texttt{Button}, the Android framework calls the \texttt{onClick()} API related to that \texttt{Button}. In other words, the developer must set the \texttt{onClick} attribute to \texttt{sendMessage} for the \texttt{Button} in \texttt{res/layout/activity\_main.xml}. Therefore, finding relations between elements statically is a cumbersome task.
 
 \begin{figure*}[htpb]
 	\centering
 	\small
 	\begin{subfigure}[t]{0.95\textwidth}
\begin{lstlisting}[language = Java , frame = trBL , firstnumber = last , escapeinside={(*@}{@*)}]
public class MainActivity extends AppCompactActivity {
    public static final String EXTRA_MESSAGE = "MESSAGE";	
    protected void onCreate(Bundle savedInstanceState) {
        super.onCreate(savedInstanceState);
        setContentView(R.layout.activity_main);}
    public void sendMessage(View view) {
        Intent intent = new Intent(this, DisplayMessageActivity.class);
        EditText editText = (EditText) findViewById(R.id.editText);
        String message = editText.getText().toString();
        intent.putExtra(EXTRA_MESSAGE, message);
        startActivity(intent);}}
\end{lstlisting}
 	\end{subfigure}\vspace{0.2cm}
 	\begin{subfigure}[t]{0.95\textwidth}
\begin{lstlisting}[language = Java , frame = trBL , firstnumber = last , escapeinside={(*@}{@*)}]
public class DisplayMessageActivity extends Activity {
    protected void onCreate(savedInstanceState) {
        setContentView(R.layout.activity_display_message);
        Intent intent = getIntent();
        String message = intent.getStringExtra(MainActivity.EXTRA_MESSAGE);
        TextView textView = findViewById(R.id.textView);
        textView.setText(message);}}
\end{lstlisting}
 	\end{subfigure}		
 	\caption{Source code for the running example~\cite{startingActivity}: the \texttt{sendMessage()} method is called whenever a user clicks the button} \label{fig:Motivation-Example}
 \end{figure*}

In this paper, we try to solve the problem of generating summaries for methods of event-driven programs by extracting the interactions between their elements at run-time. To this end, we used a deep neural network to generate summaries. Additionally, to capture interactions at run-time, we utilized dynamic call graphs.

The main contributions of this work are:
\begin{enumerate}
\item
We propose an approach to generate summaries for methods of event-driven programs. The proposed approach exploits deep neural networks and dynamic call graphs as the key components of the solution to produce meaningful summaries which not only address the semantics of the source code but also have a well-formed grammar.
\item
Unlike existing work, we introduce a novel technique for generating summaries that concentrates on run-time execution.
\end{enumerate}

The rest of the paper is organized as follows.
In \secref{Related-Work}, we review related work. We provide background and preliminary information in \secref{Background}.
In \secref{Proposed-Approach}, we describe our proposed approach. In \secref{Evaluation}, we evaluate the proposed approach by answering seven research questions. We assessed our deep neural network model using different evaluation metrics. Furthermore, we set up a user study to evaluate the generated summaries on real-world Android applications.
We conducted an experiment with 14 real-world Android applications and 42 participants to measure the quality of our approach. The experimental results show 32.3\% BLEU4 and 11.2\% METEOR.
Next, \secref{Threats-To-The-Validity} presents threats to the validity of our results. Finally, we conclude this paper and present potential future work in \secref{Conclusion}.

\section{Related Work}\label{sec:Related-Work}
In this section, we review three previous types of approaches to code summarization, namely information retrieval, machine learning, and crowdsourcing. In recent years, neural networks have been used as a new path to source code summarization. As for evaluation, BLEU4 and METEOR has been recently favored over precision and recall measures. The Java programming language is the most popular language used in source code summarization techniques.
We present an overview of these approaches in the following.

\subsection{Code Summarization via Information Retrieval}\label{sec:Information-Retrieval}
Sridhara et al.~\cite{ASESridharaHMPV2010}, proposed an algorithm for automatic description of Java methods. They preprocessed Java methods using the Software Word Usage Model (SWUM). 
SWUM is a technique for displaying methods of a program in the form of noun, verb, and adverb groups.
McBurney et al.~\cite{IEEEMcBurney2016}, introduced an approach for the automatic generation of documents for Java methods based on the context. 
In summary, this approach uses PageRank~\cite{LPage1999} to find the most important methods for the given context. SWUM helps determine what these most important methods do. Finally, natural language generation system generates a human-readable summary.
Rodeghero et al.~\cite{ICSERodeghero2014} proposed a method for choosing essential words of a code segment. They analyzed developers' eye-movements and their focused attention while writing summaries for a method, and then used their findings to weight the words subsequently. 
Antoniol et al.~\cite{IEEEAntoniolCCLM2002} proposed an approach for improving trace-ability links between a code segment and its document. They utilized the unigram language model and Vector Space Model (VSM). 
Moreno et al.~\cite{ICPCMorenoASMPV2013,ICPCMorenoMPV2013_2} presented an approach for summarizing Java classes. They primarily focused on each class content and tasks but did not heed the connection between classes. They first found a class's stereotypes and its methods. Then classified stereotypes into 13 groups. Afterward, using natural language rules, they generated a summary for each class based on a specific format. 

\subsection{Code Summarization via Machine Learning}\label{sec:Machine-Learning}
McBurney et al.~\cite{ICPCMcBurneyLMW2014}, presented a code summarization method using hierarchical topic modeling.
The most abstract description of program's tasks is given in the highest level of the hierarchy in a Hierarchical Document Topic Model (HDTM) algorithm. As one goes down the hierarchy, descriptions become more precise and clear. Authors first formed the call graph with methods as nodes and caller-callee among methods as edges of the graph. Then, HDTM	was performed on the graph.
Haiduc et al.~\cite{IEEEHaiducAMM2010}, considered code as text and exploited previous text summarization methods for summarizing code snippets as well. They used the VSM and Latent Semantic Indexing (LSI)~\cite{DPlandauer1998} in their work. 
Eddy et al.~\cite{ICPCEddyRKC2013}, proposed a code summarization algorithm using hierarchical topic modeling. In fact, this study is a replica of the Haiduc et al. work~\cite{IEEEHaiducAMM2010}, with the distinction that they utilized HPAM instead of VSM and LSI.
Programming tools help developers hide or reveal some parts of their code. This feature is known as code folding. Fowkes et al.~\cite{IEEEFowkesCRALS2017}, introduced an approach for code summarization using code folding and an Abstract Syntax Trees (ASTs).

Recently, researchers have been using neural networks as a method for generating summaries. 
Iyer et al.~\cite{Iyer-2016}, proposed CODE-NN, a neural attention model for summarizing source code.
They used LSTM networks for generating descriptions of C\# code snippets and SQL queries.
Allamanis et al.~\cite{ICMLAllamanisPS16}, introduced a novel attention mechanism using Convolutional Neural Networks (CNN)~\cite{ICMLCollobertW08}. Their goal was to generate the name of a method from its code.
Vaswani et al.~\cite{Vaswani-2017}, were the first to introduce the Transformer. Their model exploits the multi-head attention mechanism while removing recurrence and convolutions from the network. The Transformer is parallelizable and requires significantly less training time.
Hu et al.~\cite{ICPChu2018deep}, produced descriptions for Java methods using a sequence-to-sequence model. 
To improve performance, they exploited the structured form of code and introduced a novel method named SBT to parse ASTs. 
Allamanis et al.~\cite{Allamanis-2018}, represented programs with graphs to exploit the syntactic and semantic structure of source code using Gated Graph Neural Networks (GGNN). They evaluated their models based on two tasks of predicting variable names (VarNaming) and predicting variable misuse (VarMisuse).
Wan et al.~\cite{Wan2018} exploited ASTs and sequential data of code snippets in a reinforcement learning framework. The next word is predicted using the actor network and the critical component of the network evaluates the reward value.
CODE2SEQ, proposed by Alon et al.~\cite{Alon-2019}, uses syntactic structure of source code and represents each code snippet as a set of paths in an AST. Using attention mechanism of a sequence-to-sequence model, it selects relevant paths in the decoder.
LeClair et al.~\cite{LeClair-2019} also used a sequence-based model, but they treated each input data source separately. This helped the model learn the structure of code independent of other textual information in the code snippet.
Haque et al.~\cite{Haque-2020} claim the information inside a subroutine is not sufficient for summarizing it. Therefore, they proposed a sequence-based model using the attention mechanism to predict the context of subroutines.
\subsection{Code Summarization via CrowdSourcing}\label{sec:CrowdSourcing}
Badihi et al.~\cite{IEEEBadihi2017}, proposed a code summarization model for Java language using the power of crowdsourcing. They built a web-based system for developers and encouraged them to write summaries for various methods using gamification techniques. Then they collected these summaries and analyzed them to identify the most significant parts of methods from developers’ point of view.
Nazar et al.~\cite{FCSNazarJGZLR2016}, presented a code-by-code summarization approach using crowd source knowledge and supervised learning. First, they extracted code snippets from the most frequently asked questions (FAQ) section of Integrated Development Environment (IDE). Then they used four developers for labeling these code segments. 
They extracted 21 features. Then, they utilized Support Vector Machine (SVM) and Na\"ive Bayes algorithms to classify results, and finally generated summaries using these two supervised learning algorithms.
Guerrouj  et al.~\cite{ICSEGuerroujBR2015}, used the context available in posts of Stack Overflow Q\&A website in order to generate code summaries. Using an Island parser, they extracted identifiers from discussions about an element. 
Rahman et al.~\cite{SCAMRahmanRK2015}, proposed an approach to generate summaries to recommend to developers through analyzing discussions and comments of users on Stack Overflow posts. 
Wong et al.~\cite{SANERWongLT2015}, introduced a method for automatic documentation of codes using Github and clone detection techniques.

As reviewed above, there are many approaches to code summarization. However, they have their limitations. One major defect of existing solutions is that to the best of our knowledge they do not consider dynamic interactions among elements of a software program. As interactions are triggered at run-time, they cannot be inferred statically. Therefore, to exploit this information to generate better summaries for code snippets, one needs to investigate these codes at run-time. In this work, we utilize these valuable interactions to generate more useful summaries.

Another frequent shortcoming of the existing approaches is related to their evaluations. Most of the current models are evaluated using precision and recall metrics. As shown in \secref{Evaluation}, these metrics lack the validity for evaluating machine translations tasks. That is why we have used BLEU and METEOR to better evaluate the performance of our proposed model.

Moreover, most of these approaches are template-based, that is they generate summaries based on predefined rules. Therefore, these summaries neglect the essential semantics of a task/code, which renders them not very useful for end users in real-world cases. In this work, we have used deep learning methods to overcome this issue and generate more meaningful summaries.

\section{Background and Terminology}\label{sec:Background}

In this section, we provide preliminary information on the notation and methods we have used in our proposed approach.
Recently, researchers has turned to applying deep learning methods to various fields of software engineering such as commit message generation~\cite{ASEJiangAM2017,ASELiuXHLXW2018}, intention mining~\cite{IEEETSE2018_Xia}, and code search~\cite{ICSEGuZ018}. Among these fields, is code summarization via deep learning~\cite{ICPChu2018deep}, which has attained promising results so far.
The deep neural network is used as a pre-built model to generate final summaries.  The notation of our deep model is as follows\footnote{We followed the notation described at the \textit{\url{deeplearning.ai}} video tutorial~\cite{deeplearningnotation}.}:
\begin{itemize}
\item
$x$: set of source codes written in Java programming languages.
\item
$x^{(i)}$: $i$th source code in the set of $x$.
\item
$ x^{(i)}_{t} $: $t$th token in the above sequence.
\item
$ T_{s} $: $T$ is the length of the sequence $s$.
\item
$y$: set of comments written in natural language.
\item
$y^{(i)}$: $i$th comment in the set of $y$.
\item
$y^{(i)}_{t}$: $t$th term in the above sequence.
\end{itemize}

\subsection{Sequence-to-sequence Model}

Recurrent Neural Network (RNN) is suitable for sequences of inputs~\cite{werbos1990backpropagation}. RNN generates sequence $ y = (y_{1}, y_{2}, \dots, y_{T_{y}}) $ from input sequence $x=(x_{1}, x_{2}, \dots, x_{T_{x}})$.
Recently, the sequence-to-sequence model has yielded valuable results in the neural machine translation~\cite{EMNLPChoMGBBSB2014}.
In the traditional sequence-to-sequence model~\cite{NIPSSutskeverVL2014}, the decoder uses the last hidden state of the encoder as an input for generating the output sequence.
The decoder generates summaries in a deep neural network.
We aim at finding the sequence $y=(y_{1}, y_{2}, \dots, y_{T_{y}})$ from the sequence $x=(x_{1}, x_{2}, \dots, x_{T_{x}}) $, given that it applies in equation \eqref{eq:Decode}:
\begin{equation}\label{eq:Decode}
\operatorname*{arg\,max}_{y} = \prod_{t=1}^{T_{y}} P\left ( y_{t} | x, y_{1}, y_{2}, \dots, y_{t-1} \right )
\end{equation}

Suppose there are $l$ layers, and $ \overrightarrow{h}_{t}^{[l]} $ and $ \overleftarrow{h}_{t}^{[l]}  $ are the forward and backward states of the RNN for the term $t$. Therefore, $ h_{t}^{[l]} $ is computed as $ h_{t}^{[l]} = [\overrightarrow{h}_{t}^{[l]},\overleftarrow{h}_{t}^{[l]}] $. We defined a context matrix denoted as $C$. $ C_{i} $ is the $i$th column of the context matrix and is called a context vector. $ C_{i} $ indicates how much attention  the output term $ y_{i} $ pays to the terms of input sequence $ x = (x_{1}, x_{2}, \dots, x_{T_{x}}) $.
In other words, each of the members of $ x = (x_{1}, x_{2}, \dots, x_{T_{x}}) $ to what extent contributes to generating the output $ y_{i} $. States of the RNN of the decoder element are denoted as $ S_{i} $. $ \alpha_{ij} $ represents how much should $ y_{i} $ of the decoder element (the $i$th term of the summary) pay attention to $ h_{j}^{[l]} $ of the decoder element (the $j$th term of the code).
\begin{equation}\label{eq:Attention-Context}
C_{i} = \sum_{j=1}^{T_{x}} \alpha_{ij} h_{j}^{[l]}
\end{equation}

\begin{equation}\label{eq:Attention-Weights}
\begin{split}
&\alpha_{ij} = \frac{\exp\left(e_{ij}\right)}{\sum_{k=1}^{T_{x}}\exp\left(e_{ik}\right)} \\
&e_{ij} = \tanh\left(W_{j}h_{j}^{[l]} + U_{s}S_{i-1}\right)
\end{split}
\end{equation}

\subsection{PageRank Algorithm}
The PageRank algorithm was developed by Page and Brin~\cite{LPage1999} to sort webpages based on their popularity in Google's search engine. McBurney et al.~\cite{IEEEMcBurney2016} used the same concept for measuring the importance of different methods.

Damping factor, denoted as $d$, indicates how likely is it for a specific node to be visited through time. It is conventional to set $d=0.85$~\cite{BrinP1998}. The PageRank algorithm assigns a rank to each node. These ranks determine the importance of nodes in the corresponding graph. Ranks are calculated using equation \eqref{eq:pagerank}~\cite{LPage1999} which $r_{i}$ shows the rank of $n_{i}$:
\begin{equation}\label{eq:pagerank}
r_{i} = \left(1 - d\right) + d \times \sum_{n_{j} \in B_{i}} \frac{r_{j}}{l_{j}}
\end{equation}
In equation \eqref{eq:pagerank}, $l_{j}$ is the number of outgoing edges from $n_{j}$ and $B_{i}$ is the set of nodes which have outgoing edges to $n_{i}$.

\section{Proposed Approach}\label{sec:Proposed-Approach}
In this section, we present our approach to generate summaries for methods of event-driven programs. We consider the \texttt{sendMessage()} method described in \secref{introduction} as a running example. This running example is used throughout this paper to show our process of generating summaries.

\begin{figure*}[t]
	\centering
	\includegraphics[width=1.0\linewidth]{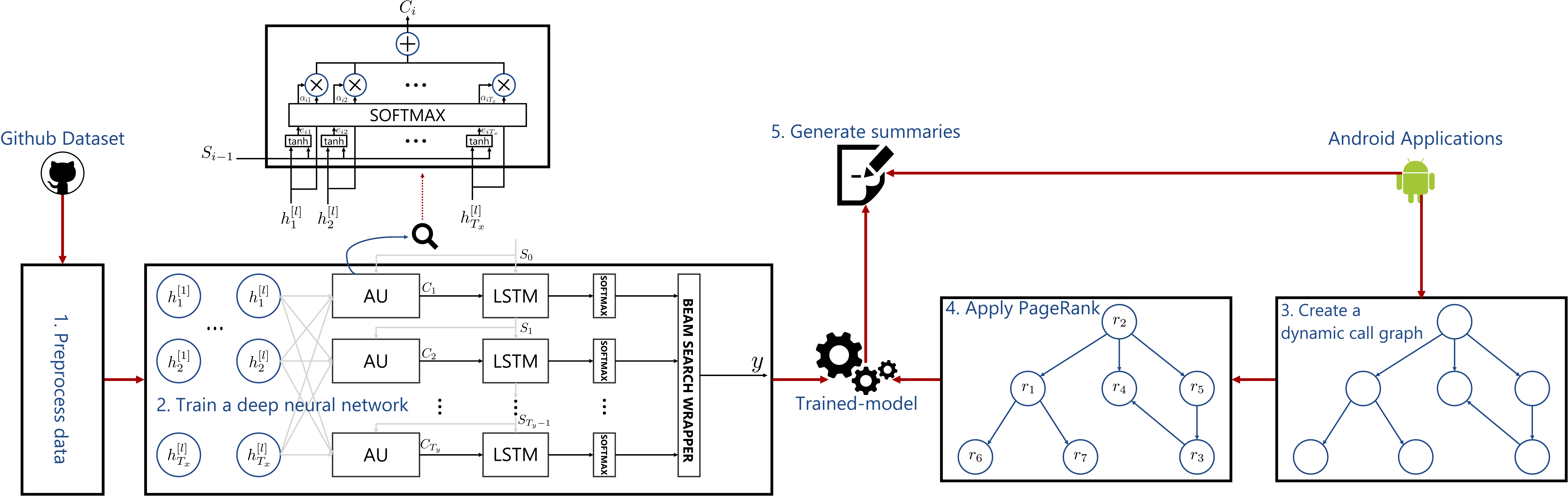}	
	\caption{An overview of the proposed approach for Android applications}\label{fig:Generating-Comments-Process}
\end{figure*}

As shown in \figref{Generating-Comments-Process}, our proposed approach consists of five steps. In the first step, we used a dataset of comment/code pairs from the Github repository. Then, applied a few preprocessing tasks on the data such as deleting blank lines, removing code snippets without summaries, and refining code-words based on the Java naming convention. 

In the second step, we built a deep neural network for the comment/code pairs. This model was used to generate the code summaries.
{\color{black}The architecture of our proposed deep neural network consists of three components, namely \textit{encoder}, \textit{decoder}, and \textit{attention mechanism}. This kind of network is mainly used for sequential data in summarization and machine translation tasks.}
We include the encoder element for encoding source code and the decoder element for decoding the encoded output to summaries. Moreover, we exploit the attention layer to put more emphasis on more important parts of the data as is frequently applied in the NLP field, specifically for machine translation and summarization tasks.

In the third step, we constructed a dynamic call graph of Android applications which were selected to generate summaries for.
In the fourth step, the PageRank algorithm was applied to the graph mentioned above to sort the methods of an application and better understand the context of a given method.

We use PageRank to identify the most important information for generating comments. 
Statically finding relations among methods is not a trivial task. However, during run-time, each block (callable code snippets or methods) is identifiable through using dynamic call graphs.
By analyzing the dynamic flow of event-driven programs, through using dynamic call graphs and then ranking them using the PageRank, we can capture long-range dependencies inside the source code and between code elements. 
Consequently, we can exploit the context and find the higher purpose behind a code snippet more accurately compared to when we just consider the method. 

In the end, using our deep neural network of step two and outputs of the PageRank algorithm, we generated human-readable summaries for the selected methods of applications. In the following, we will elaborate more on the steps of our approach.

\subsection{Step1: Preprocess Data}
In this part, we elucidate the first step of our approach. First, we performed preprocessing steps on comment/code pairs extracted from Github.
Hu et al.~\cite{ICPChu2018deep}, extracted more than 500 thousand comment/code pairs from Github and applied a few heuristic methods to extract 69708 pairs from this data. Although the 500 thousand pairs are available online, the preprocessed data are not accessible. As a result, we explored their raw data as a starting point.
These source codes are written in Java, and Java programs follow specific naming conventions. The main preprocess steps used in this study are:
\begin{enumerate}
\item
First, the blank lines (\textbackslash $\mathrm{n}$) and tabular characters (\textbackslash $\mathrm{t}$) were removed and replaced by space character.
\item
Afterward, we identified and tokenized words with all capital letters that came before words that had capital first-letters. For instance, The following regular expression does the above task:

\begin{minipage}{\linewidth}
	\small	
\begin{lstlisting}[language = Java , frame = trBL , firstnumber = last , escapeinside={(*@}{@*)}]
//SQLDatabase --> SQL Database
//Regular Expression: [A-Z]+(?=[A-Z][a-z])
\end{lstlisting}
\end{minipage}

\item
Furthermore, words with capital first-letters or all lowercase letters are extracted as well. The corresponding regular expression comes as follows:

\begin{minipage}{0.95\linewidth}
\small	
\begin{lstlisting}[language = Java , frame = trBL , firstnumber = last , escapeinside={(*@}{@*)}]
//Regular Expression: [A-Z]?[a-z]+
\end{lstlisting}
\end{minipage}
\item
Finally, we extracted words that all their letters were capital. We also kept special tokens (e.g., curly brackets and parentheses)  in the final preprocessed data.
\end{enumerate}

\figref{Running-Example-After-Stage1} demonstrates the output of our running example after the first step.

\subsection{Step2: Train a Deep Neural Network}
Our deep neural network tries to translate
$x^{(i)} = \left(x^{(i)}_{1}, x^{(i)}_{2}, ..., x^{(i)}_{T_{x}}\right)$
to
$ y^{(i)} = \left(y^{(i)}_{1}, y^{(i)}_{2}, ..., y^{(i)}_{T_{y}}\right) $ for every $i$ in a comment/code pair. {\color{black} In the following, we describe each component of the architecture in detail.}

\subsubsection{Encoder, Decoder, and Attention Mechanism}
There have been many pieces of research on the semantic representation of terms in a vector format with real numbers, namely Continuous Bag of Words (CBOW)~\cite{Mikolova2013}, SKIP-GRAM~\cite{MikolovYZ2013, GoldbergL2014, MikolovSCCD2013}, and Global Vectors (GLOVE)~\cite{EMNLPPenningtonSM2014}. The benefit of this approach is that as much as the terms are semantically similar, their vectors are similar as well.
Therefore, we used one embedding layer in the encoder and decoder components, for which the weights are tuned during the deep neural network learning phase. However, to reduce the learning time and to obtain more accurate weights, we used the pre-built model introduced in the GloVe website~\cite{glove}.

\textit{Dropout} is a simple solution to avoid overfitting~\cite{SrivastavaHKSS2014}.
Dropout randomly omits neural network units.
 We used $\mathrm{dropout}=0.2$ in the neural network layers similar to Luong et al. research~\cite{EMNLPLuongPM15}. 

Vanishing gradient is a problem in simple RNNs~\cite{hochreiter1998vanishing}.
It happens when a gradient is very small. This hinders changing values of weights and stops the neural network’s training.
 To solve this issue, various methods such as Gated Recurrent Unit (GRU)~\cite{ChungGCB2014} and Long Short-Term Memory (LSTM)~\cite{hochreiter1997long} have been proposed. We applied the latter in this work similar to Luong et al. study~\cite{EMNLPLuongPM15}. 
 
Unidirectional RNNs use only past data. However, knowing about the future helps as well. Bidirectional RNNs (BRNN) process sequences on both directions and two different layers~\cite{SchusterP97}. Graves and Schmidhuber~\cite{GravesS05}, combined bidirectional recurrent neural networks with LSTM. Moreover, one can stack layers of neural networks to build a deep  network~\cite{GravesMH13}. We have used a stack of BRNNs on top of the embedding layer in the encoder element.

When generating summaries in the decoder, one approach is to test all possible cases, which is definitely costly, with the computational complexity of $ O(|V|^{T_{y}}) $ ($ |V| $ denotes the vocabularies' set size). Another approach is to use a greedy search algorithm. These algorithms select a term that maximizes the value of $ P(y_{t} | x, y_{1}, y_{2}, \dots, y_{t-1}) $ in each step. However, if one utilizes a greedy search algorithm, she cannot change the term in the future.
Furthermore, greedy search algorithms do not guarantee to produce good results since the co-occurrence probability of some terms is higher than others.

A better solution is to exploit the beam search algorithm~\cite{GravesSeT2012}. In the beam search algorithm, the $ |B| $ top probabilities, are recorded partially for every step. $ |B| $ denotes as the width of the beam.
This heuristic algorithm does not necessarily optimize results; however, its computational complexity equals to $ O(|B| \times |V|) $ which is immensely faster than computing all cases. It is worth mentioning that if $ |B| = 1 $, this heuristic algorithm acts like a greedy one. As $ |B| $ increases, the quality of generated summaries improves, however, the learning time rises as well.

We have used the attention mechanism introduced by Bahdanau et al.~\cite{BahdanauCB2014}.
The next step is to train a model on the preprocessed data, which will be used to generate summaries in the final step.

\begin{description}[style=unboxed,leftmargin=0cm]
\item[Implementation details]
 for the deep neural network are discussed in the following. The operational environment for the deep neural network was the Ubuntu16.04. Our hardware included 40 processing cores and 64GB RAM. We used Tensorflow library to build the neural network model~\cite{OSDIAbadiBCCDDDGIIK16} and Pandas library to preprocess the data~\cite{mckinney2011pandas}.
	Embedding layers included vectors with dimensions of 300. If the input word already exists in the pre-built model, its weights from the model are used. Otherwise, the corresponding vector to the input word is initialized with uniformly distributed real numbers between -1 and 1. The pre-built model contains about 2.2 million words.
	
	The maximum length of summaries and codes are set to 35 and 100 tokens, respectively. In cases where the length of the input code is less than 100, remaining elements of the vector are replaced with zeros. Moreover, four words are pre-allocated namely $(\mathrm{PADDING}, 0)$, $(\mathrm{UNK}, 1)$, $(\mathrm{SOS}, 2)$, and $(\mathrm{EOS}, 3)$. $\mathrm{UNK} $ refers to an unknown word, which means the word does not exist in the deep neural network dictionary. Furthermore, $\mathrm{SOS}$ and $\mathrm{EOS}$ represent the start and end of each sentence, respectively.
	
	In the learning process of deep neural networks, the goal is to minimize the loss function. We have used cross-entropy loss function in this study. The set of generated summaries is represented by $(y^{(1)}, y^{(2)}, \dots, y^{(N)}) $, in which $N$ is the number of generated summaries and $ y^{(i)} $ denotes as the $i$th generated summary. Therefore, the cross-entropy loss function can be calculated using equation \eqref{eq:Loss-Function}:
	\begin{equation}\label{eq:Loss-Function}
	\mathcal{L} = \sum_{i=1}^{N} \sum_{t=1}^{T_{y^{(i)}}} -\log P\left(y^{(i)}_{t} | y^{(i)}_{1},y^{(i)}_{2}, \dots, y^{(i)}_{t-1}\right)
	\end{equation}
	
	Exploding gradient is one of the problems with long sequences. To prevent this, given that the gradient's size is more than a specific threshold such as $ \tau = 5 $, one should decrease its value using the approach introduced by Pascanu et al.~\cite{ICMLPascanuMB2013}. In other words, its value should be updated using equation \eqref{eq:Clip-Gradient}:
	
	\begin{equation}\label{eq:Clip-Gradient}
	\widehat{g} \leftarrow \frac{\tau}{\Vert \widehat{g} \Vert} \widehat{g}
	\end{equation}
	
	Adam is used for parameters' optimization~\cite{KingmaB2014}. As suggested by Kingma and Ba study~\cite{KingmaB2014}, we used $ \beta_{1} = 0.9 $, $ \beta_{2} = 0.999 $, and $ \epsilon = 10^{-8} $ as default input values for the Adam algorithm.
	
	 Overfitting is another problem that may occur while applying machine learning techniques. It happens when a technique matches too closely with a specific set of data. Therefore, rendering the aforesaid technique unfits for predicting other data sets reliably. To avoid overfitting, we randomly split the deep model's inputs into three categories, namely train ($80\%$), valid ($10\%$), and test ($10\% $) sets.
	
	 We generated checkpoints for every epochs during the training phase. Then, the best model for the validation set was selected (in terms of the BLEU score) to evaluate on the final test set.
\end{description}

\subsection{Step3: Create a Dynamic Call Graph}
\begin{figure}[ht]
	\centering
	\includegraphics[width=0.7\linewidth]{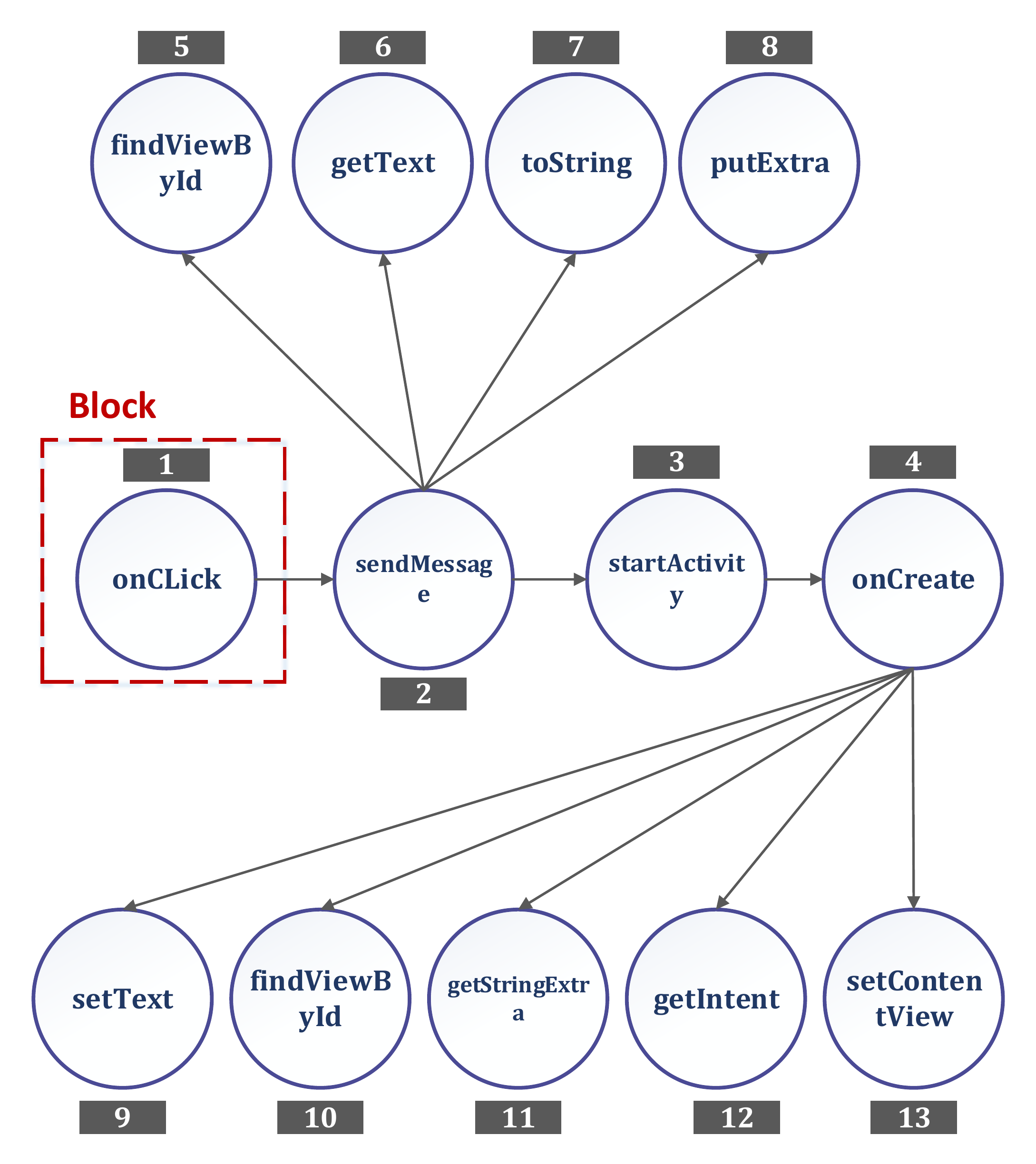}
	\caption{Call graph of the running example}\label{fig:Running-Example-Call-Graph}
\end{figure}

Event-driven programs depend on the occurrence of events at run-time. Consider the running example illustrated in \secref{introduction}. The method \texttt{sendMessage()} is invoked every time a user clicks the \texttt{Button}. However, it is not a trivial task to find out why pushing the button is followed by  running the \texttt{sendMessage()} method. In this step, we tackle this problem by leveraging the power of dynamic call graphs. Yuan et al.~\cite{FSEYuanXXPZ2017}, proposed an approach to generate a dynamic call graph for Android applications. Authors created a tool named Rundroid to create these graphs. This tool not only considers invocations between methods in static time but also recognizes messages transferred between an application and the Android framework at run-time. However, the Rundroid lacks automation, and users have to manually run and test programs to generate call graphs. Therefore, to automatize this task through generating random test with desired time intervals, we used a tool developed by Google, known as Monkey~\cite{monkey}. \figref{Running-Example-Call-Graph} depicts a part of the dynamic call graph generated for the running example.

\subsection{Step4: Apply PageRank}
We applied the PageRank algorithm to the dynamic call graph generated in the previous step.
Consider the dynamic call graph of \figref{Running-Example-Call-Graph}. This graph consists of 13 nodes denoted as $V = \left\{n_{1}, n_{2}, n_{3}, \dots, n_{13} \right\}$ and 12 edges. 
\tabref{PageRank} shows normalized results of the PageRank algorithm applied to the running example.

 \begin{table}[htpb]
 	\renewcommand{\arraystretch}{1.3}
	\centering
	\caption{PageRank Results for the Running Example}
	\label{tab:PageRank}
	\begin{tabular}{|c |c |c|}
		\hline
		\boldmath$ l_{i}$&\boldmath$B_{i}$&\boldmath$r_{i}$\\
		\hline
		\hline
		$l_{1} = 1$&$B_{1} = \{\}$&$0.0545$\\
		\hline
		$l_{2} = 4$&$B_{2} = \{1\}$&$0.1009$\\
		\hline
		$l_{3} = 1$&$B_{3} = \{2\}$&$0.0717$\\
		\hline
		$l_{4} = 5$&$B_{4} = \{3\}$&$0.1155$\\
		\hline
		$l_{5} = 0$&$B_{5} = \{2\}$&$0.0717$\\
		\hline
		$l_{6} = 0$&$B_{6} = \{2\}$&$0.0717$\\
		\hline
		$l_{7} = 0$&$B_{7} = \{2\}$&$0.0717$\\
		\hline
		$l_{8} = 0$&$B_{8} = \{2\}$&$0.0717$\\
		\hline
		$l_{9} = 0$&$B_{9} = \{4\}$&$0.0742$\\
		\hline
		$l_{10} = 0$&$B_{10} = \{4\}$&$0.0742$\\
		\hline
		$l_{11} = 0$&$B_{11} = \{4\}$&$0.0742$\\
		\hline
		$l_{12} = 0$&$B_{12} = \{4\}$&$0.0742$\\
		\hline
		$l_{13} = 0$&$B_{13} = \{4\}$&$0.0742$\\
		\hline
	\end{tabular}
\end{table}

\subsection{Step5: Generate Summaries}\label{subsec_step5}
In this step, final summaries are generated from the pre-trained model and ranks of nodes in the dynamic call graph produced in the second and fourth steps, respectively. 

We first extract methods of the selected application.
Suppose\texttt{sendMessage()} is one of these methods. First, we applied preprocessing tasks to the \texttt{sendMessage()} method. \figref{Running-Example-After-Stage1} illustrates the output of this step. Then, by using the pre-trained model from the second step, a summary was produced for the preprocessed running example (\figref{Running-Example-After-Stage2}). From the nodes in the dynamic call graph that have outgoing edges to the selected node (method), we selected the node with the highest rank.
In case of a tie, we randomly chose one of them.
We call this node a \textit{block}. If the block has a corresponding method in the source code of the program, we use that source code as an input for the pre-trained model. Otherwise, the block is related to the Android framework. In this case, we create a \textit{dummy} method by adding a signature to the block. For instance, in \figref{Running-Example-Call-Graph}, there is only one node called \texttt{onClick()}. Since the \texttt{onClick()} is related to the Android framework, we created a dummy method for the \texttt{onClick()} element and passed it to the pre-trained model (\figref{Block-as-an-Input}). 
As an example, we add ``$Public + void + onClick + (  + View + view + ) + \{ + \}$" as a dummy method for \texttt{onClick()}
After adding the output of the latter step, the summary for the given method was generated (\figref{Final-Summaries}).

Note that we do not take the Android framework’s implementation as the dummy method because of following reasons. First, all the implementations are not open-sourced. Second, implementations are not necessarily purely in Java. Third, it is possible to have multiple implementations for a given block in the Android framework, but finding the appropriate implementation can be very complicated and time-consuming. Therefore, we simply add a signature and build a dummy method. Signature addition is employed to change the format of inputs of our neural network to be like a method.

\begin{figure*}[t]
	\centering
	\begin{subfigure}[t]{0.45\textwidth}
\begin{lstlisting}[language = Java , frame = trBL , firstnumber = last , escapeinside={(*@}{@*)}]
//public void send message ( view view )
//{ ...start activity ( intent ) ; }
public void sendMessage(View view) {
  ...
  startActivity(intent);}
\end{lstlisting}
		\caption{Output of the first step on the running example}
		\label{fig:Running-Example-After-Stage1}
	\end{subfigure}	
	\hfil	
	\begin{subfigure}[t]{0.45\textwidth}
\begin{lstlisting}[language = Java , frame = trBL , firstnumber = last , escapeinside={(*@}{@*)}]
//Sends a message to the specified service
public void sendMessage(View view) {
  ...
  startActivity(intent);}
\end{lstlisting}
		\caption{Output of the second step on the running example}	
		\label{fig:Running-Example-After-Stage2}
	\end{subfigure}	
	\hfil
	\begin{subfigure}[b]{0.45\textwidth}
\begin{lstlisting}[language = Java , frame = trBL , firstnumber = last , escapeinside={(*@}{@*)}]
//Called whenever a view has been clicked
public void onClick(View view) {}
\end{lstlisting}
		\caption{Pass the highest node block as an input to the pre-trained model}	
		\label{fig:Block-as-an-Input}
	\end{subfigure}
	\hfil
	\begin{subfigure}[b]{0.45\textwidth}
\begin{lstlisting}[language = Java , frame = trBL , firstnumber = last , escapeinside={(*@}{@*)}]
//Sends a message to the specified service
//Called whenever a view has been clicked
public void sendMessage(View view) {
  ...
  startActivity(intent);}
\end{lstlisting}
		\caption{Generated summaries for the given method}	
		\label{fig:Final-Summaries}
	\end{subfigure}
	\hfil
	\caption{The outputs of applying different steps of the proposed approach on the running example}
\end{figure*}

\section{Evaluations}\label{sec:Evaluation}
In this section, we present the results of both qualitative and quantitative evaluation of our proposed approach. First, the {\color{black} pre-trained} deep neural network was assessed using BLEU4 and METEOR metrics. Then, using an empirical study, we examined the usefulness of our approach in aiding developers understand the objective of methods. This qualitative assessment was performed on 14 Android applications and 42 methods, using three highly skilled experts for generating reference summaries and 42 Android developers for evaluating generated summaries.

\subsection{Evaluations of Deep Neural Network}
We first present Research Questions (RQs), evaluation metrics and the evaluation process. Afterward, we discuss results of our evaluations and analyze them subsequently.

\subsubsection{Research Questions}
To evaluate our deep neural network, we answer the following four questions:
\begin{description}[style=unboxed,leftmargin=0cm]
	\item[RQ1:]
	How much the proposed model has been successful in learning comments/codes sets?
	\item[RQ2:]
	What is the precision of the generated summaries by the proposed model?
	\item[RQ3:]
	What proportion of reference summaries were retrieved as the final generated summaries?
	\item[RQ4:]
	How well has the proposed deep neural network performed compared to the other baseline deep neural networks?
\end{description}

\subsubsection{Evaluation Metrics}
Here, we investigate the evaluation metrics used in this study, namely BiLingual Evaluation Understudy (BLEU) and Metric for Evaluation of Translation with Explicit
ORdering (METEOR).
\begin{description}[style=unboxed,leftmargin=0cm]
\item[BLEU]
is used for automated evaluation of machine translation algorithms~\cite{AMACLPapineni2002}. Since code summarization is a type of translation of code snippets to human-readable summaries, BLEU can be used for evaluating abstractive code summarization. {\color{black}
BLUE refines precision by valuing each term exactly for as many times as it has appeared in the reference translations. $P_1$ considers every term separately. $P_2$ has a similar concept as $P_1$, except that it computes the precision of bigrams.
}
We compute the precision for different values of $n$ in n-grams. The final score is calculated using equation \eqref{eq:BLEU-SCORE}. In this equation, $ BP $ is a penalty for short summaries, which are identified using equation \eqref{eq:BP-Cal}. In equation  \eqref{eq:BP-Cal},  $ r $ and $ c $ are the lengths of reference and generated summaries, respectively.
\begin{equation}\label{eq:BLEU-SCORE}
\mathrm{BLEU} = BP \times \exp\left(\frac{\sum_{i=1}^{N} \log(P_{i})}{N}\right)
\end{equation}

\begin{equation}\label{eq:BP-Cal}
BP = \begin{cases}1 \qquad c > r \\\exp\left(1 - \frac{r}{c}\right) \qquad c \le r \end{cases}
\end{equation}

\item[METEOR]
was proposed to mitigate BLEU's shortcomings~\cite{ACLBanerjeeL2005}. METEOR focuses mainly on recall, unlike BLEU which pays more attention to precision. METEOR is based on the term-to-term mapping of the generated summary with its corresponding reference summary. 

METEOR is calculated using equation \eqref{eq:Meteor}. In equation \eqref{eq:Meteor}, $ R $ is the refined recall, $ P $ is the refined precision and $ PN $ is the penalty (which is issued for having only unigrams).
\begin{equation}\label{eq:Meteor}
\mathrm{METEOR} = \frac{10RP}{R + P} \times \left(1 - PN\right)
\end{equation}

In equation \eqref{eq:PN-Cal}, $C$ is the number of common chunks between the generated and reference summaries.
In cases that the reference and generated summaries are identical, there is only one chunk. On the other hand, if there exists only uni-grams, number of chunks equals to the number of term-to-term mappings denoted as $M_{u}$.

\begin{equation}\label{eq:PN-Cal}
PN = 0.5 \times \left(\frac{C}{M_{u}}\right)^{3}
\end{equation}

\end{description}
\subsubsection{Evaluations Setup}
Our deep neural network uses Github {\color{black}data} as an input resource.
Some of the pairs did not have enough comment length. Therefore, we removed pairs with comments shorter than four words. Furthermore, some of the comments are too long to be used for training deep neural networks.
Ying et al.~\cite{FSEYingR2014}, claimed that most of the summaries are less than three sentences. Moreno et al.~\cite{ICPCMorenoASMPV2013}, stipulated that summaries with less than 20 terms are suitable for comment generation. Consequently, comments with more than 35 words were removed from the pairs. Similarly, source codes with more than 100 tokens were removed. Some of the comments neither were written in English nor were produced by a human. We removed these automatically generated comments as well. Finally, by applying a few minor heuristics (e.g., converting tokens to lowercase), we selected 71257 pairs of comment/code. \tabref{Datasets-stats} describes statistical information about these pairs.

\begin{table}[H]
	\renewcommand{\arraystretch}{1.3}
	\centering
	\caption{Statistical Information of Extracted Pairs}
	\label{tab:Datasets-stats}
	\begin{tabular}{l c c c c c}
		\hline
		&\bfseries Mean&\bfseries Q1&\bfseries Q2&\bfseries Q3&\bfseries \# Unique tokens\\
		\hline
		\hline
		Comment length&$11.25$&$7$&$9$&$14$&$44934$\\
		\hline
		Code length&$33.49$&$15$&$26$&$47$&$22525$\\
		\hline
	\end{tabular}
\end{table}

\subsubsection{Evaluations Results}
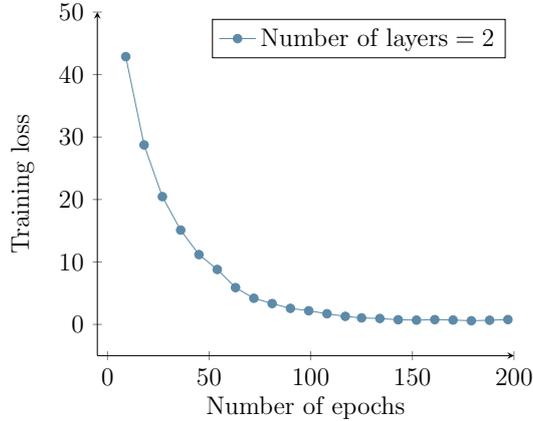
\begin{figure}[h]
	\centering
	\begin{tikzpicture}[scale=0.8]
	\begin{axis}[
	axis lines=left,
	xmin=-5, xmax=200,
	ymin=-5, ymax=50,
	xlabel={Number of epochs},
	ylabel={Training loss},
	grid style=dashed
	]

		\addplot[color=airforceblue, mark=*] coordinates {
		(9,42.87)
		(18,28.74)
		(27,20.46)
		(36,15.10)
		(45,11.18)
		(54,8.80)
		(63,5.89)
		(72,4.19)
		(81,3.34)
		(90,2.57)
		(99,2.19)
		(108,1.71)
		(117,1.29)
		(125,1.04)
		(134,0.95)
		(143,0.75)
		(152,0.69)
		(161,0.77)
		(170,0.70)
		(179,0.59)
		(188,0.67)
		(197,0.78)
	};

	\legend{$\text{Number of layers}=2$}
	\end{axis}
	\end{tikzpicture}
	\caption{Cross-entropy loss values based on different epochs.}
	\label{fig:Training-Loss}
\end{figure}

To answer the RQ1, we used \textit{perplexity} metric~\cite{BrownPPLM1992}. Perplexity estimates how well a deep neural network can perform on a training dataset. It is calculated using $ \mathrm{Perplexity} = \exp(\mathcal{L}) $, in which $ \mathcal{L} $ is the cross-entropy loss function.
\tabref{BLEU-METEOR-RUN} shows the best perplexity values in the 10 last epochs. Moreover, \figref{Training-Loss} presents cross-entropy loss function based on different epochs.

To answer the RQ2, we used the BLEU4 metric. The maximum number of terms for generated summaries is $35$, which is considered short. Therefore, based on the suggestion of Papineni et al.~\cite{AMACLPapineni2002}, we used the maximum four-grams in calculating the value of this metric. \tabref{BLEU-METEOR-RUN} illustrates BLEU4 results based on different parameters.
We set the number of epochs, batch size, and beam width in all cases to 200, 512, and 50, respectively. We achieved the best BLEU4 score, $31.4$, using a network of two layers, employing both pre-trained embedding layer and SBT.

To answer the RQ3, we used METEOR metric. 
We achieved the best METEOR score, $13.1$, using a network with the same properties mentioned above. \tabref{BLEU-METEOR-RUN} presents the values of discussed metrics for different parameters.

  \begin{table}[h]
 	\renewcommand{\arraystretch}{1.3}
	\centering
	\caption{Results of the Proposed Deep Neural Network with Different Parameters}
	\label{tab:BLEU-METEOR-RUN}
	\begin{tabular}{|c|c|c |c|c|c|}
		\hline
		\rotatebox{90}{\bfseries \# of layers}&\rotatebox{90}{\bfseries \twoelementtable{Pre-trained}{embedding layer}}&\rotatebox{90}{\bfseries {\color{black} SBT}}&\rotatebox{90}{\bfseries BLEU4}&\rotatebox{90}{\bfseries METEOR}&\rotatebox{90}{\bfseries Perplexity}\\
		\hline
		\hline
		$4$&\cmark&{\color{black}\xmark}&$30.8$&$11.3$&$3.53$\\
		\hline
		$4$&\xmark&{\color{black}\xmark}&$31.0$&$6.7$&$5.16$\\
		\hline
		$3$&\cmark&{\color{black}\xmark}&$30.7$&$11.8$&$1.95$\\
		\hline
		$2$&\cmark&{\color{black}\xmark}&$30.9$&$12.7$&$\boldsymbol{1.84}$\\
		\hline
		{\color{black}$2$}&{\color{black}\xmark}&{\color{black}\cmark}&{\color{black}$ 30.3 $}&{\color{black}$7.8$}&{\color{black}$5.47$}\\
		\hline
		{\color{black}$2$}&{\color{black}\cmark}&{\color{black}\cmark}&{\color{black}$\boldsymbol{31.4}$}&{\color{black}$\boldsymbol{13.1}$}&{\color{black}$2.18$}\\
		\hline
	\end{tabular}
\end{table}

To answer the RQ4, we used six code summarization techniques proposed by
Haque et al.~\cite{Haque-2020}, LeClair et al.~\cite{LeClair-2019}, Alon et al.~\cite{Alon-2019}, Allamanis et al.~\cite{Allamanis-2018}, Vaswani et al.~\cite{Vaswani-2017}, and Iyer et al.~\cite{Iyer-2016} as our baselines. We selected these approaches because they are the state-of-the-art in the field of code summarization ranging from traditional sequence-based neural networks such as the RNN to the Transformer. Furthermore, they have been published in the leading venues of both SE and NLP.
We used the publicly available implementation provided by Haque et al.~\cite{Haque-2020}.
\tabref{DeepNeuralNetwork-Evaluation} presents results of the proposed deep neural network's performance in compared with the baselines.

\begin{table}[!h]
 	\renewcommand{\arraystretch}{1.3}
	\centering
	\caption{Comparing the proposed model to the baselines}
	\label{tab:DeepNeuralNetwork-Evaluation}
	\begin{tabular}{|c|c|c|}
		\hline
		\bfseries Approach&{\bfseries METEOR}&{\bfseries BLEU4}\\
		\hline
		Iyer et al., 2016~\cite{Iyer-2016}              &   $18.4$                &   $20.0$    \\
		\hline
		Vaswani et al., 2017~\cite{Vaswani-2017}        &   $8.0$                &   $7.8$    \\
		\hline
		Allamanis et al., 2018~\cite{Allamanis-2018}    &   $18.3$   &   $20.5$    \\
		\hline
		Alon et al., 2019~\cite{Alon-2019}              &   $\boldsymbol{19.1}$                &   $21.0$    \\
		\hline
		LeClair et al., 2019~\cite{LeClair-2019}        &   $18.5$                &   $20.5$    \\
		\hline
		Haque et al., 2020~\cite{Haque-2020}            &   $18.9$                &   $21.3$    \\
		\hline
	    Proposed approach                               &   $13.1$               &   $\boldsymbol{31.4}$    \\
		\hline
	\end{tabular}
\end{table}

\subsubsection{Quantitative Analysis of Results}
According to \tabref{BLEU-METEOR-RUN}, cross-entropy loss function has decreased $2.18$ per word in perplexity. This indicates that the proposed model has efficiently performed on the training dataset. Furthermore, according to \figref{Training-Loss}, the model has not progressed significantly after epoch number 170. Therefore, we believe increasing the epoch number to more than $200$ does not improve the performance of the model very much.
It is worth mentioning that our proposed model does not necessarily compete with Hu et al.~\cite{ICPChu2018deep}. In fact, our model can complement their approach. To demonstrate this, we applied SBT to our model and obtained higher scores in terms of BLEU4 (31.4) comparing to the previous result without SBT (30.9).
\tabref{DeepNeuralNetwork-Evaluation} compares the proposed neural network with the state-of-the-art.
Our model significantly improved BLEU4 comparing to those approaches.
In particular, comparing to Haque et al.~\cite{Haque-2020}, our model improved BLEU4 from $21.3$ to $31.4$. However, except from Vaswani et al.~\cite{Vaswani-2017}, other approaches have better METEOR than our model ($13.1$ compared to $19.1$).

\subsection{Evaluations of Generated Summaries}
Here, we evaluate the usefulness of our model in aiding Android applications' comprehension using an empirical study.
\subsubsection{Research Questions}
To evaluate generated summaries, we investigate the following questions:
\begin{description}[style=unboxed,leftmargin=0cm]
	\item[RQ1]
	Considering the reference summaries, how accurate have been the generated summaries?
	\item[RQ2]
	How well has the proposed model performed compared to other approaches?
	\item[RQ3]
	How good is the quality of the generated summaries?
\end{description}
\subsubsection{Evaluations Setup}
We used a dataset of 14 open-source Android applications with different sizes and features from previous studies ~\cite{ICSEMirzaeiGBSM2016,QRSDengOS17} to evaluate generated summaries.
\tabref{Android-Applications} presents some information about these applications.

\begin{table}[h]
	\renewcommand{\arraystretch}{1.3}
	\centering
	\caption{Set of Android Applications Used in the Study}
	\label{tab:Android-Applications}
	\begin{tabular}{|c|c|c|c|}
		\hline
		\bfseries Application Name&\bfseries \# of lines&\bfseries \# of Methods&\bfseries \# of Classes\\
		\hline
		\hline
		Tister&$ 423$&$14$&$8$\\
		\hline
		Hashpass&$ 429$&$8$&$2$\\
		\hline
		Munchlife&$631$&$17$&$9$\\
		\hline
		Justsit&$849$&$43$&$13$\\
		\hline
		Blinkenlightsbattery&$851$&$61$&$14$\\
		\hline
		Autoanswer&$999$&$50$&$13$\\
		\hline
		Anycut&$1095$&$60$&$18$\\
		\hline
		Dofcalculator&$1321$&$14$&$9$\\
		\hline
		Divideandconquer&$1824$&$156$&$28$\\
		\hline
		Passwordmakerpro&$2824$&$282$&$67$\\
		\hline
		TippyTipper&$2953$&$148$&$36$\\
		\hline
		Tokenlist&$3680$&$225$&$43$\\
		\hline
		Httpmon&$4939$&$392$&$86$\\
		\hline
		Remembeer&$5915$&$257$&$54$\\
		\hline
	\end{tabular}
\end{table}

\begin{figure}[!h]
	\centerline{\includegraphics[width=0.7\linewidth]{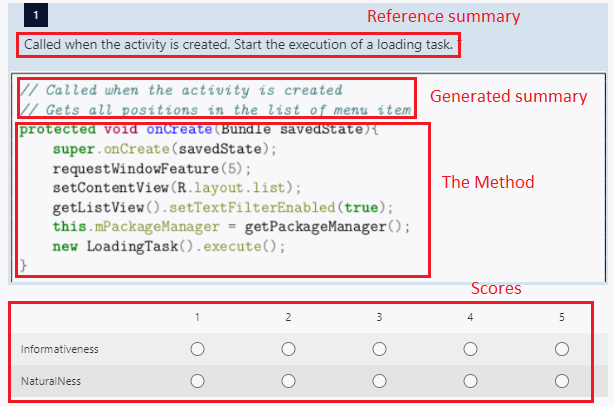}}
	\caption{A sample question from the questionnaire}
	\label{fig:sample_question}
\end{figure}

First, we randomly selected {\color{black}three} methods from each application and 
{\color{black}two experts who are not the authors of this paper, worked with the 14 applications and investigated the related parts of the source code to know more about the context of each selected method. Then, wrote summaries for the 42 methods independently. After that, they compared their comments in a session and chose a common comment for each method. In case of disagreement, they asked a third expert for his opinion and they all came to an agreement for the comments of these methods together. It is worth mentioning that the two experts' average Android programming experience was 5.5 years and the third expert was a senior Android developer with 8 years of experience. Also, there was about 18\% of disagreement between the first two experts, which all were resolved with the help of the third expert.}
Afterward, using the proposed model and the baseline approaches, we generated candidate summaries for each method. 
Note that our baselines are not specifically designed for event-driven programs, so the comparison may not be completely fair. However, these are the closest approaches to our problem that we are aware of so far. 
We designed an online questionnaire to qualitatively assess the performance of these models. Each question in the questionnaire contains (1) a method, (2) its reference summary, and (3) a generated summary by one of the baselines.
\figref{sample_question} presents a question from the questionnaire. 
We asked the participants to score the generated summary based on the method itself and the reference summary provided to them.
The method along with its reference summary help the participants obtain a better understanding of each method and its context. Note that by knowing what exactly each method does, participants were stricter in evaluating the generated summaries. This helped us remove the chance of careless or wrong scoring. Note that the generated summaries were presented in random orders, to remove any bias from the experiment. That is participants did not know which comment was written by which approach.
As both the number of methods and baselines in our study are high, we could not evaluate each method with all the six baselines plus our approach using all participants. This would require a very long questionnaire which subsequently would result in a decline in the accuracy of evaluations. Therefore, we randomly assigned the questions to them and made sure to cover each of the 42 methods at least by 7 participants for each baseline.

The 42 participants were majored in computer science, with an average of 7.6 years of general programming experience, 5.8 years of Java programming experience, and 4.3 years of Android programming experience.
It took each participant on average 73 minutes to finish the questionnaire.
We analyzed the generated summaries from two perspectives; their \textit{informativeness} and \textit{naturalness}~\cite{Iyer-2016};
\begin{description}[style=unboxed,leftmargin=0cm]
\item[Informativeness:] What proportion of the important parts of the code does the generated summary cover.
\item[Naturalness:] How smooth and human-readable is the generated summary. Naturalness also takes into account the syntax of each sentence.
\end{description}
Participants scored the generated summaries for each method based on a 1-5 star scaling. Description of each score is as follows:

\begin{itemize}
	\item
	Informativeness:
	\begin{enumerate}
		\item
		The output does not describe the method's functionality to any extent.
        \item
        Only insignificant parts of the code are covered in the summary.
        \item
        Some important parts of the code are covered in the summary.
        \item
        Most of the important parts are covered in the summary.
        \item
        All significant and essential parts of the code are well summarized.
    \end{enumerate}
    \item
    Naturalness:
    \begin{enumerate}
        \item
        The output is not readable by humans at all.
        \item
        The output is barely understandable with many errors.
        \item
        The output is understandable but has noticeable syntax errors.
        \item
        The output is understandable but has negligible syntactical errors.
        \item
        The output is completely understandable with no syntactical errors.
	\end{enumerate}
\end{itemize}

\subsubsection{Evaluations Results}
To answer the RQ1, we calculated BLEU4 and METEOR metrics for each method. We compared our new results to the Iyer et al.~\cite{Iyer-2016}, Vaswani et al.~\cite{Vaswani-2017}, Allamanis et al.~\cite{Allamanis-2018}, Alon et al.~\cite{Alon-2019}, LeClair et al.~\cite{LeClair-2019}, and Haque et al.~\cite{Haque-2020} approaches to answer the RQ2. Tables~\ref{tab:Approach-Baseline-Comparision} presents the comparison results of these  approaches.

 \begin{table}[!h]
 	\renewcommand{\arraystretch}{1.3}
	\centering
	\caption{Comparing the proposed model to the baselines}
	\label{tab:Approach-Baseline-Comparision}
	\begin{tabular}{|c|c|c|}
		\hline
		\bfseries Approach&{                           \bfseries BLEU4}         &{\bfseries METEOR}\\
		\hline
		Iyer et al., 2016~\cite{Iyer-2016}            &   $11.4$              &   $17.9$  \\
		\hline
		Vaswani et al., 2017~\cite{Vaswani-2017}      &   $8.6$              &   $9.7$   \\
		\hline
		Allamanis et al., 2018~\cite{Allamanis-2018}  &   $13.1$              &   $\boldsymbol{18.0}$   \\
		\hline
		Alon et al., 2019~\cite{Alon-2019}            &   $11.1$              &   $16.4$   \\
		\hline
		LeClair et al., 2019~\cite{LeClair-2019}      &   $12.0$              &   $16.9$   \\
		\hline
		Haque et al., 2020~\cite{Haque-2020}          &   $12.1$              &   $16.0$   \\
		\hline
	    Proposed approach                             &   $\boldsymbol{32.3}$ &   $11.2$   \\
		\hline
	\end{tabular}
\end{table}

\begin{figure}[htpb]
	\centerline{\includegraphics[width=0.9\linewidth]{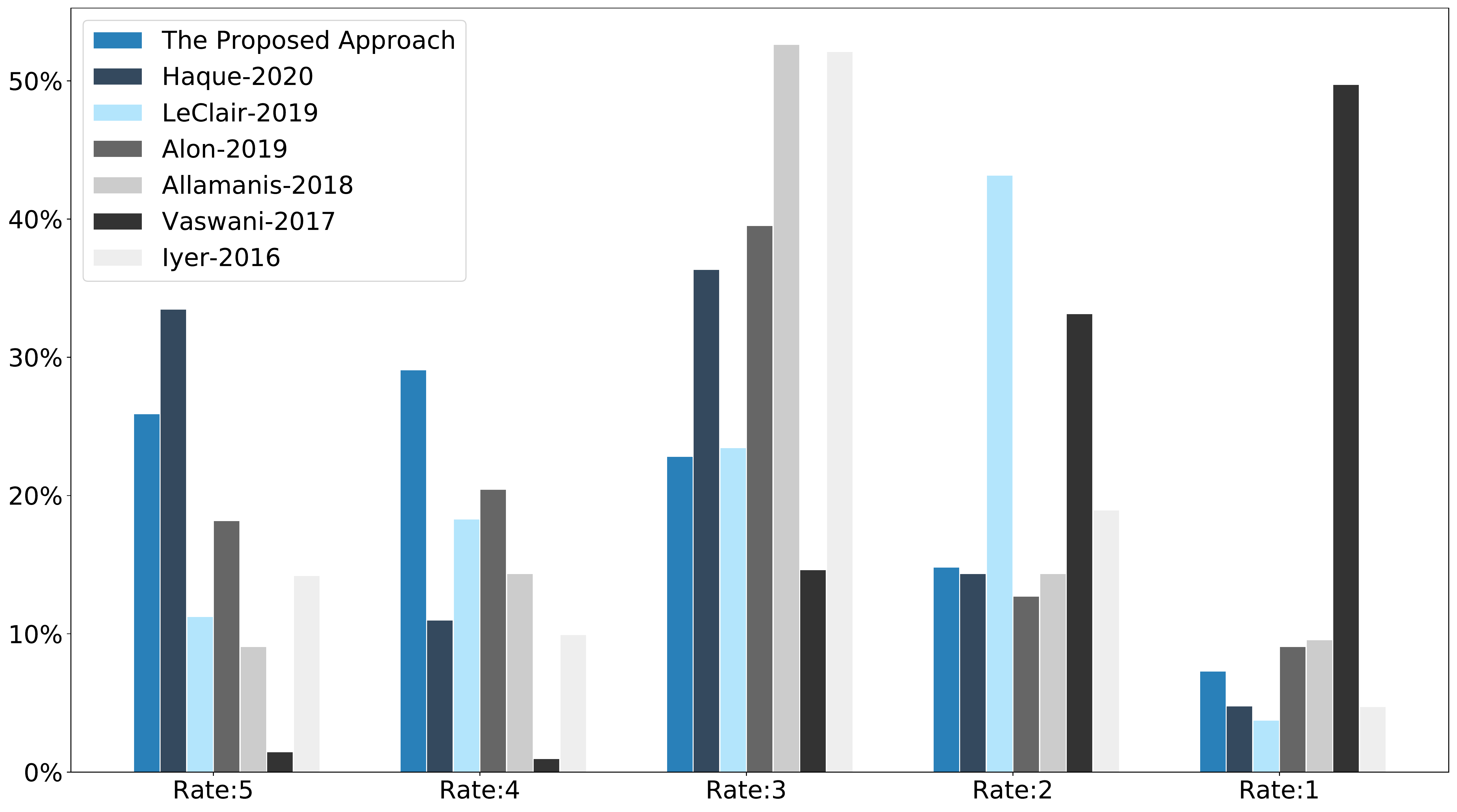}}
	\caption{Distribution of informativeness}
	\label{fig:Iformativeness-Dist}
\end{figure}

\begin{figure}[htpb]
	\centerline{\includegraphics[width=0.9\linewidth]{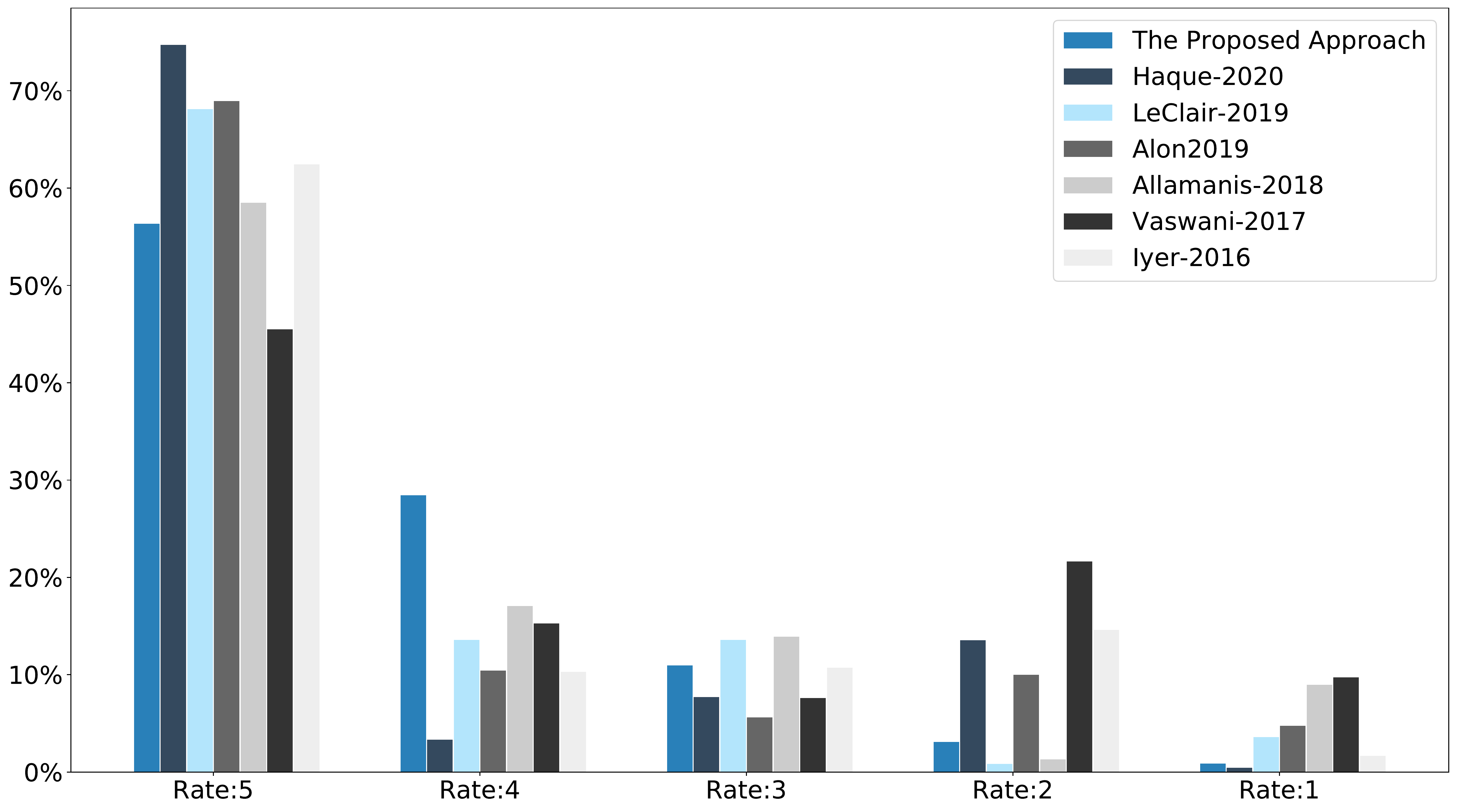}}
	\caption{Distribution of Naturalness}
	\label{fig:Naturalness-Dist}
\end{figure}

To answer the RQ3, \figref{Iformativeness-Dist} and \figref{Naturalness-Dist} present the distribution of informativeness and naturalness variables. The mean scores for informativeness and naturalness of our proposed approach for all participants are $3.51$ and $4.36$.
The results indicate that if one only considers rating 5 (perfect summaries) and rating 4 (good enough summaries with negligible mistakes) as desirable outcomes of a summarization method, our approach outperforms all the others regarding both informativeness and naturalness.

\subsubsection{Quantitative Analysis of Results}\label{subsec:Quantitative_Analysis}

According to \tabref{Approach-Baseline-Comparision}, average BLEU4 and METEOR of the proposed approach are  $32.3$ and $11.2$, respectively.
To investigate whether there is a significant difference between the results of our proposed approach and other existing approaches, we 
followed the guideline and the tool provided by Herbold~\cite{Herbold2020}.
We conducted a statistical analysis for 7 approaches with 42 samples.
We used the non-parametric Friedman test to investigate difference between the median values of the approaches~\cite{friedman1940}. We employed the post-hoc Nemenyi test to determine which aforementioned differences are statistically significant~\cite{nemenyi1962}. The Nemenyi test used \textit{critical distance} (CD) to evaluate which one is significant. If the difference is greater than CD, then the two approaches are statistically significant different.

\begin{figure}[htpb]
	\centerline{\includegraphics[width=0.9\linewidth]{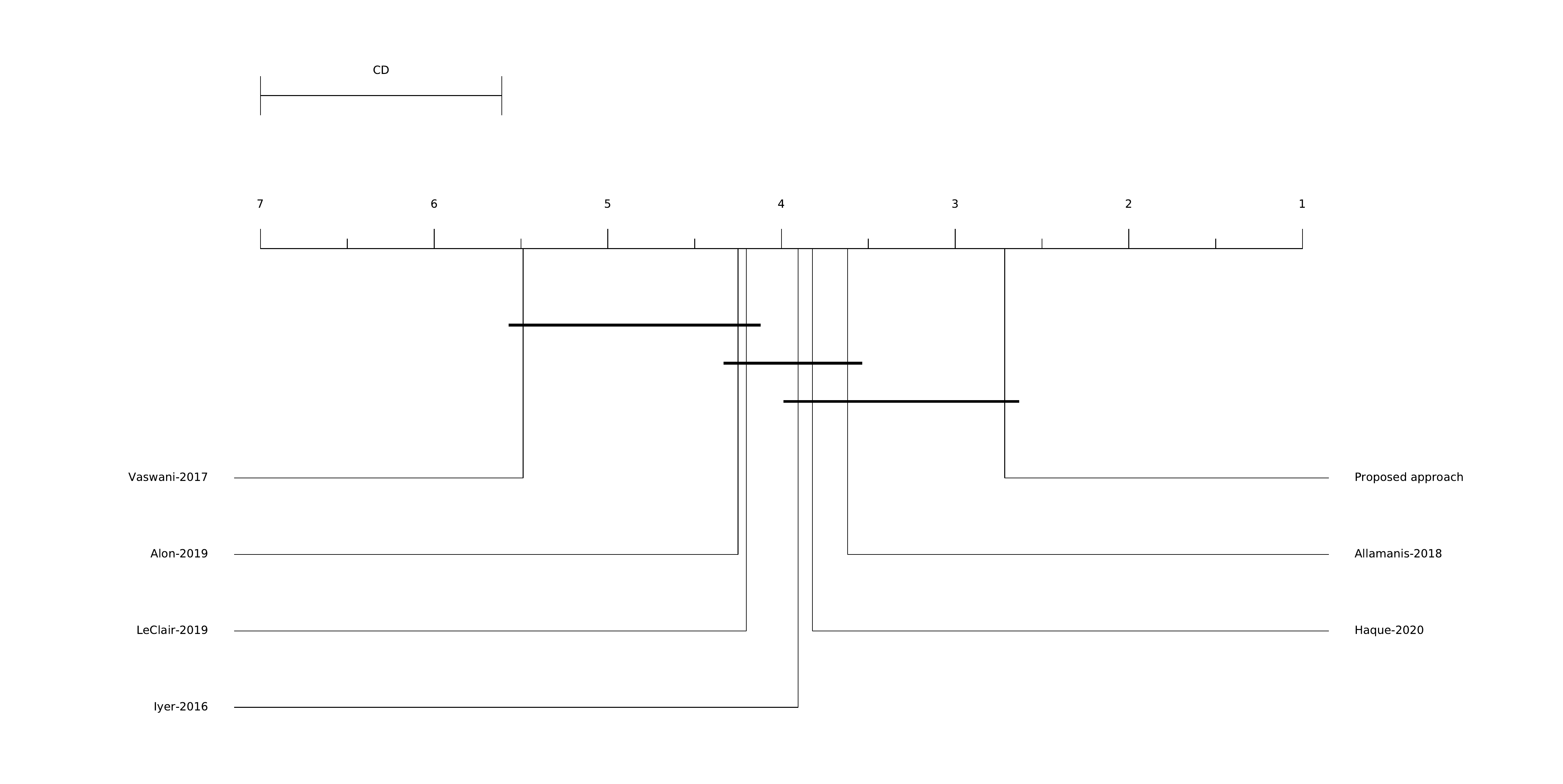}}
	\caption{The results of hypothesis testing for BLEU4}
	\label{fig:hyp-blue}
\end{figure}

\figref{hyp-blue} depicts the results of tests for BLEU4.
The Friedman test rejects the null hypothesis that there is no difference between median values of the approaches. Consequently, we accept the alternative hypothesis that there is a difference between the approaches. Based on the \figref{hyp-blue} and the post-hoc Nemenyi test, we cannot say that there are significant differences within the following approahces: (the proposed approach, Allamanis et al.~\cite{Allamanis-2018}, Haque et al.~\cite{Haque-2020} and Iyer et al.~\cite{Iyer-2016}); (Allamanis et al.~\cite{Allamanis-2018}, Haque et al.~\cite{Haque-2020}, Iyer et al.~\cite{Iyer-2016}, LeClair et al.~\cite{LeClair-2019}, and Alon et al.~\cite{Alon-2019}); (LeClair et al.~\cite{LeClair-2019}, Alon et al.~\cite{Alon-2019}, and Vaswani et al.~\cite{Vaswani-2017}). All of the other differences are statistically significant.

\begin{figure}[htpb]
	\centerline{\includegraphics[width=0.9\linewidth]{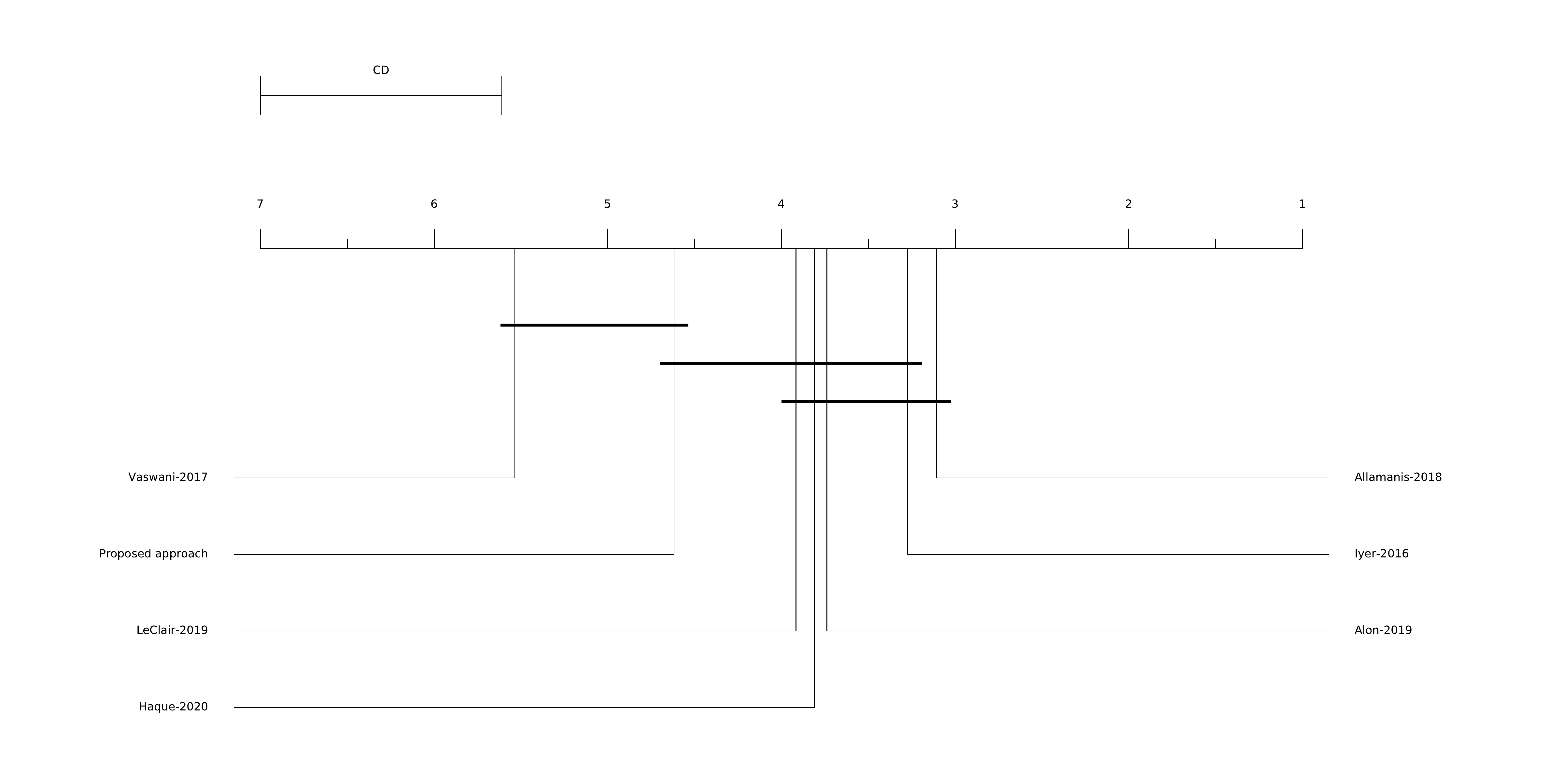}}
	\caption{The results of hypothesis testing for METEOR}
	\label{fig:hyp-meteor}
\end{figure}

\figref{hyp-meteor} shows the results of tests for METEOR. The Friedman test rejects the null hypothesis that there is no difference between median values of the approaches. Consequently, we accept the alternative hypothesis that there is a difference between the approaches. Based on the 
\figref{hyp-meteor} and the post-hoc Nemenyi test, we cannot say there are significant differences within the following approaches: (Allamanis et al.~\cite{Allamanis-2018}, Iyer et al.~\cite{Iyer-2016}, Alon et al.~\cite{Alon-2019}, Haque et al.~\cite{Haque-2020}, and LeClair et al.~\cite{LeClair-2019}); (Iyer et al.~\cite{Iyer-2016}, Alon et al.\cite{Alon-2019}, Haque et al.~\cite{Haque-2020}, LeClair ~\cite{LeClair-2019}, and the proposed approach); (the proposed approach and Vaswani et al.~\cite{Vaswani-2017}). All of the other differences are statistically significant.

{\color{black} Sample means for informativeness and naturalness of the proposed approach  are $\overline{X}_{\mathrm{inf}} = 3.51$ and $\overline{X}_{\mathrm{nat}} = 4.36$}, respectively. We applied t-distribution to estimate mean and standard deviation of informativeness and naturalness results.
Using equation \eqref{eq:sigmahat}, and the confidence level of 95\% ($\alpha = 0.05$),

\begin{equation} \label{eq:sigmahat}
\begin{split}
&\hat{\sigma}_{\overline{X}_{\mathrm{inf}}} = \tfrac{S_{\mathrm{inf}}}{\sqrt{N}} = {0.19}\\
&\hat{\sigma}_{\overline{X}_{\mathrm{nat}}}= \tfrac{S_{\mathrm{nat}}}{\sqrt{N}} = {0.13}.
\end{split}
\end{equation}

We can compute equation \eqref{eq:mu},

\begin{equation} \label{eq:mu}
\mu = \overline{X} \pm t_{(N-1={\color{black}41},\alpha=0.05)}\times \hat{\sigma}_{\overline{X}}.
\end{equation}

Therefore, we can conclude that

\begin{equation} \label{eq:sigmarange}
\begin{split}
&\hat{\sigma}_{\overline{X}_{\mathrm{inf}}} = { 3.51 \pm 0.38}\\
&\hat{\sigma}_{\overline{X}_{\mathrm{nat}}} = {4.36 \pm 0.27.}
\end{split}
\end{equation}

Equation \eqref{eq:sigmarange} shows that with the confidence level of 95\%, by increasing the number of participants, in the worst case the mean scores for informativeness and naturalness will be greater than {$3.13$} and {$4.09$}, respectively.

\subsubsection{Qualitative Analysis of Results}

According to {\color{black}\figref{Iformativeness-Dist}, \figref{Naturalness-Dist}} and our definition of informativeness and naturalness metrics, we conclude as follows:

\begin{enumerate}
	\item
	Participants in	$25.9$\% of cases reported that generated summaries cover all essential parts of the codes.
	\item
	Participants in	$55.0$\% of cases reported that in the worst case, generated summaries cover many salient features of the code.
	\item
	Participants only in $22.1$\% of cases reported that generated summaries are not related to the codes or document just trivial code snippets.
	\item
	Participants in $45.0$\% of cases reported that generated summaries have neglected a few necessary parts of the codes.
	\item
	Participants in	$56.3$\% of cases reported that generated summaries are human-readable and do not have any syntactical error.
	\item
	Participants in	$84.8$\% of cases reported that in the worst case, the generated summaries have minor syntactical errors.
	\item
	Participants in	$11.0$\% of cases reported that generated summaries are human-readable but have major syntactical errors.	
	\item
	Finally, participants only in $4.1$\% of cases reported that generated summaries are barely human-readable.	
\end{enumerate}

\subsubsection{Threats to Validity of the Empirical Study}

In this section we discuss the threats to validity of our empirical study.
We evaluated the quality of final summaries extracted from 14 Android applications and 42 methods. It is reasonable that the quality of extracted methods affects results. To reduce this threat, we sampled randomly from extracted methods.
Moreover, 42 individuals performed our qualitative assessment. Therefore, the outcome of this section depends on characteristics of the individuals taking the questionnaire, such as their mood, the time it took them to fill the questionnaire, and other natural factors. To reduce this threat, we tried to have a large number of participants.
Note that for a large-scale evaluation, we need a dataset of Android methods, their comments and the APK file as the requirement for the Rundroid tool. However, to the best of our knowledge, available datasets only contain Java methods, and unfortunately, we were unable to collect such a large dataset for Android methods. It is worth mentioning that one can collect a dataset for a large-scale study; however, this is not a trivial task. But we plan to address this issue in future work.
Moreover, we investigated whether there is a significant difference between the results of our proposed approach and other existing approaches using the Friedman and post-hoc Nemenyi test. We have reported the results in Section \ref{subsec:Quantitative_Analysis}.
The results indicate that there is a significant difference between our approach and other baselines regarding performance based on BLEU and METEOR. Also, in the same section, we proved that with the confidence level of 95\%, by increasing the number of participants, in the worst case the mean scores for informativeness and naturalness of our approach will be greater than 3.13 and 4.09 (scale of 5 stars), respectively. We believe all these can attest to the good performance of our approach in a large-scale study.

As mentioned above, 42 individuals performed our qualitative assessment of the generated summaries for the selected real-world Android applications’ methods. Because the number of participants is limited, we cannot extend our results to the rest of the developers’ community. To reduce this threat, we have tried to select a well-distributed sample of developers to assist in the evaluation phase.
We also investigated whether there is a significant difference between our results and other existing approaches' results. We performed the Friedman and post-hoc Nemenyi test and reported the results.

\subsection{Discussion}\label{subsec:Discussion}

Here, we discuss the limitations of our proposed approach. 
The quality of generated comments would be restricted when a target method did not represent the characteristics of 
event-driven programs. Helper functions (e.g., math functions), utility functions (e.g., logging) and interfaces are among these methods. \figref{Example-helper-function} demonstrates a utility function which calculates the time elapsed from the start. 
This function is used in various methods to compute the elapsed time. Indeed, there is no meaningful context for this method.
Consequently, the second generated comment does not relate to the method's functionality or context.
Our proposed approach fail to generate high quality comments for this method. One way to handle this situation is to apply threshold analysis. For example, if the block rank is less than given threshold, it means that the method does not have meaningful context and one could simply ignore that block. However, in practice, determining the threshold value automatically is not a trivial task, and can be addressed in future work.

\begin{figure}[h]
	\centering
	\footnotesize
\begin{lstlisting}[language = Java , frame = trBL , firstnumber = last , escapeinside={(*@}{@*)}]
// Gen: Returns the elapsed time of states
// Gen: Called when the activity is updated
// Ref: Return how much time has passed since the starting time in milliseconds.
public long getElapsed() {
    return System.currentTimeMillis() - this.start;
}
\end{lstlisting}
\caption{A utility function extracted from the HTTPMON application}\label{fig:Example-helper-function}
\end{figure}

\section{Threats to Validity}\label{sec:Threats-To-The-Validity}
In this section, we review threats to the validity of our research findings, categorizing possible threats into four groups of internal, external, construct and dependability ones~\cite{SEKEFeldtM2010}.

\subsection{Internal Validity}
Internal validity asks whether the variables used in the proposed approach affect the outcomes and whether they are the only influential factors in the study~\cite{SEKEFeldtM2010}.

The dynamic call graph in Rundroid is constructed based on tests that are run on Android applications. These tests are run manually in the original version of the study~\cite{FSEYuanXXPZ2017}. Therefore, how the tests are run and their runtime impact results. To reduce this threat, authors generated 5000 random events using the Monkey tool to minimize human intervention in the tests.

We believe the quality of code snippets affects the quality of generated summaries as well. Logically training the models on high-quality source codes can help the model generate better summaries. However, not all real-life projects benefit from high-quality source code. Moreover, quality is a subjective concept and can be  interpreted differently in various cases. Therefore, it is not a trivial task to collect a high-quality dataset of code snippets.

In case of a tie in Section \ref{subsec_step5}, we randomly choose a node. We did not investigate the effects of this choice since they were rare cases that a tie happened. So we believe it was not necessary. But we acknowledge that different decisions in these cases can provide different outcome.

\subsection{External Validity}
External validity includes how expandable are results of a study, can they be used in other contexts, and do cause and effect relationships hold with other conditions as well~\cite{SEKEFeldtM2010}.

In this study, we have used Rundroid to build call graphs. Rundroid is developed for generating call graphs in the Android framework. Therefore, it is not suitable for usage in other event-driven programs. To reduce this threat, we plan to investigate and use other tools in near future.
In the first and second phase of the proposed approach, we have used deep neural networks. The deep neural network architecture can be employed in other contexts, namely other natural language processing fields.

Moreover, in the evaluation phase, we have only used the Android framework’s examples as event-driven programs. Thus, it is not guaranteed that our approach will perform the same on other event-driven programs such as web-based programs. In future, we plan to address this threat by evaluating our model on other event-driven platforms.

\subsection{Construct Validity}
Construct validity includes theoretical concepts and discussions of the experiment and the use of appropriate evaluation metrics~\cite{SEKEFeldtM2010}.

Theoretical concepts used in this work, have been already evaluated and proved by the academic society. The proposed approach is a combination of different methods in a new context. We have evaluated the generated summaries not only by valid and reliable quantitative metrics but also through human qualitative judgment. Results indicate that the employed approach has been successful in generating summaries.

\subsection{Dependability}
Dependability validity answers to questions such as whether the findings are compatible, and whether the experiment and its results are reproducible~\cite{SEKEFeldtM2010}.

\begin{description}[style=unboxed,leftmargin=0cm]
\item[Compatibility]
We evaluated the final generated summaries quantitatively and qualitatively. As shown in previous sections, their outcomes are compatible.
\item[Reproducibility]
We have used deep neural networks (which are inherently based on probability) to generate the summaries of event-driven programs’ methods. To reduce this threat, we have set the number of epochs to 200. This is because cross-entropy loss function is almost stable after the 170th epoch and did not decrease in our experiments. Also, the preprocessed input data is available online for other researchers at \textit{\url{https://github.com/ase-sharif/deep-code-document-pairs}}.
It is worth mentioning that we have tried our best to minimize human intervention in all steps to make results more independent and reliable.
\end{description}

\section{Conclusions and Future Work}\label{sec:Conclusion}
Code summarization is a useful technique for helping developers comprehend and maintain software programs more efficiently. There are different approaches for summarizing code segments, namely utilizing information retrieval, machine learning, and crowdsourcing knowledge. However, existing approaches do not take into account the interactions between different parts of the code while the program is running. Through exploiting deep neural networks and dynamic generation of the call graph, we tried to overcome the deficiencies of previous work and generate summaries that are more suitable. Results of the proposed approach were evaluated both qualitatively and quantitatively. We used BLEU4, METEOR, precision and recall metrics for quantitative assessment and an online questionnaire for assessing the informativeness and naturalness of generated summaries from developers’ perspectives as a means of qualitative assessment. Our results indicate that the proposed approach outperforms state-of-the-art techniques.

As for future work, one of the conventional solutions while using the sequence-to-sequence models is to employ a convolutional layer in the encoding component~\cite{EMNLPKalchbrennerB2013}. Adding this layer helps the deep neural network attain additional information about the words around each word. The use of a convolutional layer has improved results in machine translation studies. We intend to exploit this layer in future and analyze its effect on the proposed model.

Moreover, the Android framework is only one example of event-driven applications. In the future, we are going to examine other frameworks to evaluate the proposed approach and expand our findings.

\section*{Acknowledgment}
The authors would like to thank the participants who assessed the quality of our proposed approach.



\bibliographystyle{elsarticle-num}

\bibliography{ref}

\end{document}